\documentclass[12pt]{article}
\usepackage[left=1.0in, right=1.0in, top=1.0in, bottom=1.0in]{geometry}
\usepackage{authblk}    
\usepackage{abstract}   
\usepackage{mathptmx, amsfonts, bm}	
\usepackage{amsmath}
\usepackage{mathrsfs}
\usepackage{enumitem}
\usepackage{graphicx}
\usepackage{caption}
\usepackage{subcaption}
\usepackage{listings}
\usepackage{float}
\usepackage{tikz}
\usetikzlibrary{shapes,arrows}
\usepackage{xcolor}
\usepackage{todonotes}
\definecolor{navyblue}{HTML}{0072B2}
\definecolor{blue}{HTML}{0F9ED5}
\definecolor{orange}{HTML}{E69F00}

\newcommand{\blue}[1]{\textcolor{blue}{#1}}
\newcommand{\firstline}[1]{\textcolor{black}{#1}}

\usepackage{hyperref}
\hypersetup{
    colorlinks=true,    
    urlcolor=blue,
    citecolor=blue,
    linkcolor=black,
}
\usepackage{cleveref}

\title{Hybrid Physics-Data Enrichments to Represent Uncertainty in Reduced Gas-Surface Chemistry Models for Hypersonic Flight}
\date{}
\author[1]{Rileigh Bandy\thanks{Corresponding author: rjbandy@sandia.gov}}
\author[2]{Rebecca Morrison}
\author[3]{Erin Mussoni}
\author[1]{Teresa Portone}

\affil[1]{Optimization and Uncertainty Quantification Department,
Sandia National Laboratories, Albuquerque, NM}
\affil[2]{Computer Science Department,
University of Colorado, Boulder, CO}
\affil[3]{Thermal/Fluid Science and Engineering Department,
Sandia National Laboratories, Livermore, CA}

\begin{document}

\maketitle

\begin{abstract}
During hypersonic flight, air reacts with a planetary re-entry vehicle's thermal protection system (TPS), creating reaction products that deplete the TPS. Reliable assessment of TPS performance depends on accurate ablation models.
New finite-rate gas-surface chemistry models are advancing state-of-the-art in TPS ablation modeling, but model reductions that omit chemical species and reactions may be necessary in some cases for computational tractability.
This work develops hybrid physics-based and data-driven enrichments to improve the predictive capability and quantify uncertainties in such low-fidelity models while maintaining computational tractability.
We focus on discrepancies in predicted carbon monoxide production that arise because the low-fidelity model tracks only a subset of reactions.
To address this, we embed targeted enrichments into the low-fidelity model to capture the influence of omitted reactions.
Numerical results show that the hybrid enrichments significantly improve predictive accuracy while requiring the addition of only three reactions.
\end{abstract}

\section{Introduction}

\firstline{
Models in computational science and engineering simulate complex physical systems and guide design and decision-making.}
For instance, in aerospace engineering, models are used to simulate the hypersonic trajectory of a planetary re-entry vehicle and inform heat shield design.
To make such simulations computationally feasible, models often incorporate simplifying assumptions—such as reduced chemical mechanisms, neglected physical interactions, or idealized boundary conditions.
While these approximations are necessary, they can lead to model-form errors that degrade the model’s predictive accuracy.

\firstline{
The design of an effective thermal protection system (TPS) for a planetary re-entry vehicle relies accurately modeling gas-surface interactions.}
During re-entry, high heat fluxes and chemically non-equilibrium flows expose the heat shield to reactive atomic species that drive ablation--the removal of surface material through chemical and thermal reactions \cite{prata2022air}. 
Therefore, ablative carbon-based shields are commonly applied, with oxidation as a primary mechanism \cite{poovathingal2017finite}.
At extreme temperatures, gas ionization and carbon sublimation may occur, leading to reactions with atomic oxygen (O) and nitrogen (N) \cite{kline2018boundary} or the transition of solid carbon to gas \cite{zhluktov1999viscous}.
Quantifying these gas-surface reactions is essential to TPS design, as an excessively thick shield adds unnecessary weight and cost, while a shield that is too thin risks catastrophic failure.

\firstline{
Experimental testing of TPS is often prohibitively expensive, with full-scale wind tunnel tests costing roughly $\$100,000$ per day \cite{osti_1118141} and hypersonic flight experiments costing $\$11-286$ million as of 2009 \cite{berry2011recommendations}.}
Thus, design relies heavily on simulations.
New trajectories may require finite-rate gas-surface chemistry models, which are non-equilibrium models and significantly more expensive than equilibrium chemistry models \cite{prata2022air}. 
High-fidelity, non-equilibrium chemistry models offer accuracy but are too costly for design sweeps. 
Cheaper, low-fidelity models are needed, though they introduce model-form error and reduce predictive accuracy.

\firstline{
Therefore, we investigate model enrichment, where the goal is to improve an existing but deficient low-fidelity model.}
A common approach is the model discrepancy approach, which treats the low-fidelity model as a biased approximation and adds a stochastic discrepancy term to each model output \cite{kennedy2001bayesian, higdon_computer_2008}.
This method does not modify the low-fidelity model itself making it easy to apply across different models, but there are key limitations.
Extrapolation beyond the calibration regime is often unreliable \cite{maupin_model_2020}, and discrepancy terms are restricted to observed quantities so they cannot propagate to unobservable quantities of interest (QoIs).

\firstline{
To address these issues, intrusive enrichments improve a deficient low-fidelity model by directly modifying its governing equations.} 
Intrusive enrichments embed learned or physics-based terms into the model, enabling a more consistent treatment of interdependent outputs and allowing extrapolation to unobserved QoIs.
However, their effectiveness depends on how the enrichments are modeled, constrained, and trained \cite{oliver2015validating, morrison2018representing, portone2022bayesian, kaheman2022automatic, bandy2024stochastic, wu2024learning}. 
For example, in the modified Sparse Identification of Nonlinear Dynamics (modified SINDy) algorithm \cite{kaheman2022automatic}, the enrichment is selected from a library of candidate functions, it is unconstrained, and the enriched model is deterministic. 
Modified SINDy requires a well-posed problem and approximates the true system. 
However, the true dynamics may involve hundreds of terms on the right-hand side (RHS) of the differential equations, and enough data is seldom available to train them all. 

\firstline{
Intrusive enrichments span a spectrum from physics-based to data-driven.}
Physics-based enrichments, such as those in \cite{oliver2015validating}, embed domain knowledge directly into the model, but can be time-consuming to develop and require expert knowledge to build.
Data-driven approaches, like those in \cite{acquesta2022model, wu2024learning}, are more flexible when physical intuition is lacking.
However, they lack direct supervision, may be unstable for complex models, are costly in high dimensions, and struggle to generalize beyond the training input range.

\firstline{
While both intrusive and nonintrusive approaches can perform comparably in calibration settings, intrusive enrichment is generally more robust for complex models involving multiple, coupled outputs.}
By modifying the model itself, it naturally captures interactions between outputs and supports prediction of new and unobservable QoIs.

\firstline{In this study, we propose a hybrid, intrusive model enrichment framework that combines physics-based insights with a supervised data-driven enrichment to improve the accuracy of low-fidelity models.}
First, physics-based enrichments are constructed using theory from a high-fidelity model.
Then, a pointwise enrichment is used to pre-train the data-driven enrichment before it is embedded into the model, enabling supervised learning that enhances stability and reduces computational cost, particularly in high-dimensional settings.
We apply this approach to the motivating example of predicting heat shield recession.

The paper is organized as follows. In \Cref{finite-rate}, we introduce finite-rate gas-surface chemistry models, review the high-fidelity model, derive a low-fidelity model, and define our inputs and the quantity of interest. 
In \Cref{enrichment} we propose embedding physics-based and data-driven enrichments in the low-fidelity model to characterize uncertainty and reduce error caused by omitted species. 
Results are discussed in \Cref{results}, and concluding remarks are given in \Cref{conc}.
\section{Finite-Rate Gas-Surface Chemistry Modeling}\label{finite-rate}

\begin{table}[H]
\captionsetup{skip=2pt}
\caption{Nomenclature}
\caption*{\small Variables and constants defined with units (if applicable).}
    \begin{tabular}{lll}
        $A_v$ &$=$& Avogadro  constant, mol$^{-1} $ \\
        $c$ &$=$& Molar concentration, mol$\cdot$m$^{-3}$ \\
        $B$ &$=$& Total active site density, mol$\cdot$m$^{-2}$
        \\
         $E$ & $=$ & Activation energy, J\\
         $F$ & $=$ & One-fourth the mean thermal speed of a gas species, m$\cdot$s$^{-1}$\\
         $f$ & $=$ & Flux of a species, mol$\cdot$m$^{-2}\cdot$s$^{-1}$\\
         $h$ &$=$& Planck's constant, J$\cdot$s \\
         $k_b$ & $=$ & Boltzmann constant, J$\cdot$K$^{-1}$ \\
         $M$ & $=$ & Molar mass, kg$\cdot$mol$^{-1}$ \\
         $m$ & $=$ & Mass of an atom or molecule, kg \\
         $P$ & $=$ & Partial pressure, Pa\\
         $R$ & $=$ & Universal gas constant, J$\cdot$K$^{-1}\cdot$mol$^{-1}$\\
         $S$ & $=$ & Adsorption selectivity\\
         $T$ & $=$ & Temperature, K \\
         $x$ &$=$& Longitudinal distance from the stagnation point, m \\
         $\gamma$ &$=$& Pre-exponential factor \\
         $\rho$ & $=$ & Density, kg$\cdot$m$^{-3}$ \\
         $\chi$ &$=$& Mole fraction\\
        $G$ &$=$& A subscript distinguishing gas species ($G \in \left\{\text{O}, \text{N}, \text{O}_2\right\}$)
    \end{tabular}
    \label{tab:nomenclature}
\end{table}

\firstline{
Accurately predicting the molar flux of carbon monoxide (CO) is crucial for assessing carbon ablation effects, making it a primary quantity of interest (QoI) in finite-rate gas-surface chemistry models.}
In this work, we use the air-carbon ablation (ACA) model \cite{prata2022air} as our high-fidelity reference.
To improve computational efficiency, we derive a low-fidelity model by omitting certain reactions.
We demonstrate the discrepancies between the high- and low-fidelity models caused by these omissions using a simulation test representative of hypersonic environments. This section includes a description of the air-carbon ablation model, a formulation of a low-fidelity model, and a description of the inputs and QoI for our test case. Key symbols and constants used throughout this section are summarized in \Cref{tab:nomenclature}.

\subsection{Background: Air-Carbon Ablation Model}
\begin{figure}[H]
    \centering
    \begin{subfigure}[b]{0.2\textwidth}
         \centering
         \includegraphics[width=0.8\textwidth]{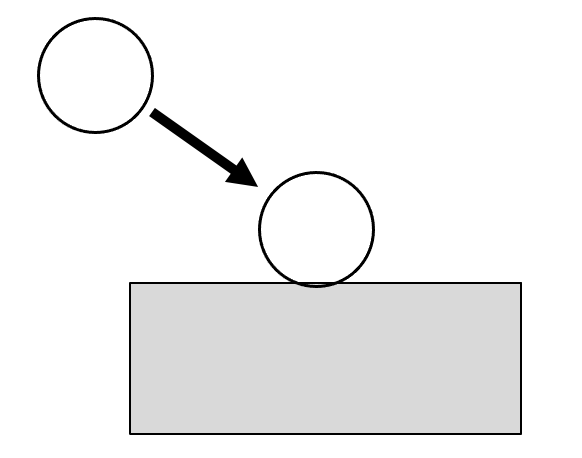}
         \caption{Adsorption}
         \label{fig:ad}
     \end{subfigure}
     \hfill 
     \begin{subfigure}[b]{0.19\textwidth}
         \centering
         \includegraphics[width=0.8\textwidth]{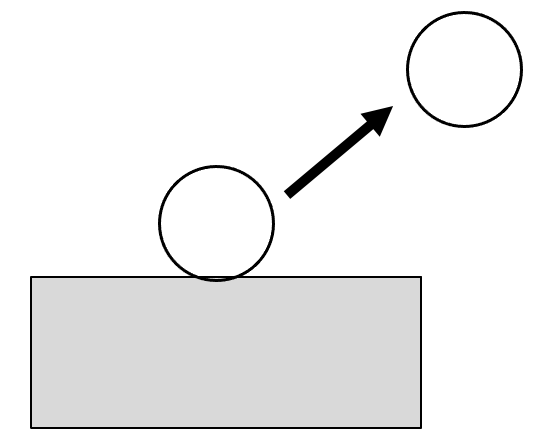}
         \caption{Desorption}
         \label{fig:des}
     \end{subfigure}
     \hfill 
     \begin{subfigure}[b]{0.19\textwidth}
         \centering
         \includegraphics[width=0.8\textwidth]{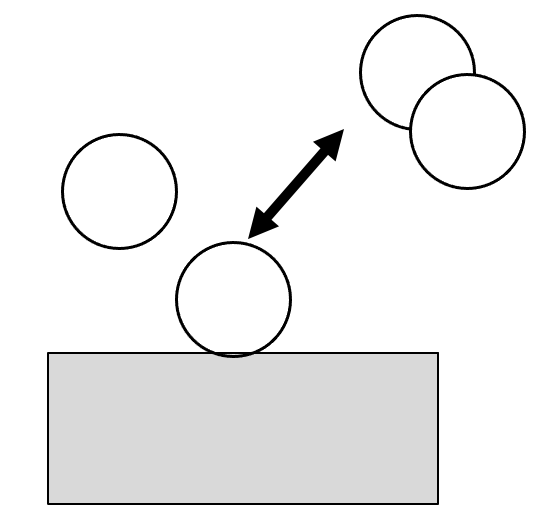}
         \caption{Eley-Rideal}
         \label{fig:er}
     \end{subfigure}
     \hfill 
     \begin{subfigure}[b]{0.28\textwidth}
         \centering
         \includegraphics[width=0.8\textwidth]{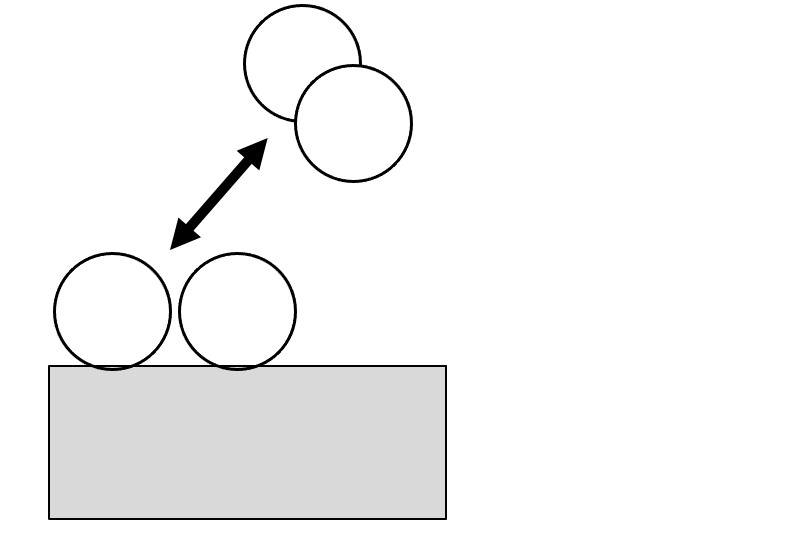}
         \caption{Langmuir-Hinshelwood}
         \label{fig:lh}
     \end{subfigure}
    \caption[Depictions of reactions in the ACA model]{The four types of gas-surface reactions in the ACA model are: (a) adsorption, (b) desorption, or recombination reactions via (c) Eley-Rideal or (d) Langmuir-Hinshelwood.}
    \label{fig:reactions}
\end{figure}

\firstline{
The air-carbon ablation (ACA) model, shown in \Cref{tab:prata}, involves $20$ reaction mechanisms that focus on the reactions between the carbon surface and gas species \cite{prata2022air}.}
The model describes oxidation by O and molecular oxygen, O$_2$; nitridation by N; and surface catalyzed recombination of both O and N.
The ACA model involves four types of reactions: adsorption, desorption, Eley-Rideal, and Langmuir-Hinshelwood, all of which are illustrated in \Cref{fig:reactions}. Reaction rates are determined by an empirical function for each reaction type.
The mechanisms associated with these reaction types are discussed below, and reaction notation is defined subsequently.

\begin{table}[H]
    \caption{ACA gas-surface chemistry model reaction set}
    \centering
    \begin{tabular}{clccl}
                \hline
                Number   & Reaction & $S$ or $\gamma$ &   $E/R$ & Type    \\\hline
                1  & O + (s)  $\rightarrow$  O(s)   & 0.3 &  0      & Adsorption \\
                2  & O(s) $\rightarrow$ O + (s) & 1.0 &  44277      & Desorption \\
                3  & O + O(s) + C(b) $\rightarrow $ CO + O + (s) & 100.0 &  4000 & Eley-Rideal \\
                4  & O + O(s) + C(b) $\rightarrow$ CO$_2$ + (s)  & 1.0 &  500 & Eley-Rideal  \\
                5  & O + (s)  $\rightarrow$ O$^*$(s)  & 0.7  &  0      & Adsorption  \\
                6  & O$^*$(s) $\rightarrow$ O + (s)  & 1.0  & 96500   & Desorption  \\
                7  & O + O$^*$(s) + C(b) $\rightarrow $ CO + O + (s)  & 1000 & 4000 & Eley-Rideal  \\
                8  & O$^*$(s) +  O$^*$(s) $\rightarrow$ O$_2$ + 2(s) & $10\textsuperscript{-3}$  & 15000 & Langmuir-Hinshelwood  \\
                9  & O(s) +  O(s) $\rightarrow$ O$_2$ + 2(s)& $5.0\times 10\textsuperscript{-5}$ &  15000  & Langmuir-Hinshelwood \\
                10  & N + (s) $\rightarrow$ N(s) & 1.0 &  2500   & Adsorption \\
                11  & N(s) $\rightarrow $ N + (s) & 1.0 &  73971 & Desorption \\
                12  & N + N(s) + C(b) $\rightarrow$ CN + N + (s)  & 1.5 &  7000 & Eley-Rideal  \\
                13  & N + N(s)  $\rightarrow$ N$_2$ + (s) & 0.5  &  2000      & Eley-Rideal \\
                14  & N(s) +  N(s) $\rightarrow$ N$_2$ + 2(s)  & 0.1  & 21000  & Langmuir-Hinshelwood  \\
                15  & N(s) + C(b) $\rightarrow$ CN + (s) & $10\textsuperscript{8}$ & 20676       & Eley-Rideal  \\
                16  & O$_2$ + 2(s) $\rightarrow$ 2O(s) & 1.0  & 8000     & Adsorption  \\
                17  & O$_2$ + O(s) + C(b) $\rightarrow$ CO + O$_2$ + (s) & 100.0 &  4000     & Eley-Rideal \\
                18  & O$_2$ + O(s) + C(b) $\rightarrow $ CO$_2$ + O + (s) & 1.0 &  500 & Eley-Rideal \\
                19  & O$_2$ + 2(s) $\rightarrow$ 2O$^*$(s)  & 1.0 &  8000 & Adsorption  \\
                20  & O$_2$ + O$^*$(s) + C(b)  $\rightarrow$ CO + O$_2$ + (s) & 1000.0  &  4000      & Eley-Rideal  \\
                \hline
    \end{tabular}
    \caption*{\textit{Note:} For Reaction 15, the rate coefficient is given by $k_{15} = 10^8\exp(-\frac{20676}{T})$.}
    \label{tab:prata}
\end{table}

\firstline{
Adsorption is the initial step in gas-surface reactions, where a gas-phase species binds to an empty surface site before undergoing further reactions.}
Adsorption reactions involve a gas-phase reactant $G$, where $G \in \left\{\text{O}, \text{N}, \text{O}_2\right\}$, and an empty surface site $(s)$, which functions like a chemical species. Using the Arrhenius law, the adsorption rate coefficient is
\begin{equation}
    k_{ad} = \frac{F_G}{B^k}S\exp{\left(
    -\frac{E}{RT}\right)},
\end{equation}
where
\begin{equation*}
    F_G = \frac{1}{4}\sqrt{\frac{8k_b T}{\pi m_G}}
\end{equation*} 
is one-fourth the mean thermal speed of the gas species (in moles per meter). 
The mass of the gas species is $m_G$. 
The power of $B$ depends on the gas species being adsorbed.
For O and N, $k=1$ and for O$_2$, $k=2$. 
The surface sites that the gas species can adsorb to are represented by $B$, the total active site density.
The adsorption selectivity (or the sticking probability) is $S$, which is commonly set to one unless the gas species could bind to the surface in multiple ways.
For example, the ACA model captures weakly-bonded and strongly-bonded O to the surface.
There is a $30\%$ probability that O will weakly bond resulting in O$(s)$, and a $70\%$ chance that O will strongly bond resulting in O$^*(s)$.
The activation energy is experimentally derived from molecular beam data \cite{prata2022air}.
Finally, the rate of adsorption is $k_{ad}[G][(s)]$, where 
$[G]$ is the flux of the gas-phase species (in mole per cubic meter per second)
and $[(s)]$ is the surface density of the empty reactive sites.

\firstline{
Desorption describes the release of adsorbed gas species from the surface, with a rate governed by temperature-dependent kinetics and surface species density.}
The adsorbed gas species desorbs from the surface with a rate coefficient of
\begin{equation}
    k_{des} = \frac{2\pi m_G k_b^2 T^2}{A_v B h^3}\exp{\left(
    -\frac{E}{RT}\right)}.
\end{equation}
The  rate of desorption is $k_{des}[G(s)]$, where $[G(s)]$ is the surface density of the adsorbed species.
The adsorbed species are $G(s) \in \left\{\text{O}(s), \text{O}^*(s), \text{N}(s) \right\}$ or weakly-bonded O, strongly-bond O, and bonded N with no distinction of bonding strength.
Note, oxygen can form O$_2$ in the air, but adsorbing to the surface would require two empty sites.
In this model, the molecule would have to split to attach to the surface.
This is theoretically true for nitrogen as well, and the ACA model allows for the production of molecular nitrogen, N$_2$.
However, it does not account for the nitridation by N$_2$, which is a simplifying assumption. 
The impact of this assumption depends on the chemical composition of the gas.
This simplification is generally valid because, although nitrogen is more abundant in the bulk atmosphere, oxygen dominates the surface chemistry in the environments considered in this work.

\firstline{
Eley-Rideal reactions describe gas-surface interactions between a gas-phase species and an adsorbed species that remove surface carbon and produce gas-phase products.}
These reactions involve one gas-phase reactant, a carbon atom in the bulk media of the surface C$(b)$, and one chemisorbed species, which is a gas species adsorbed to the surface.
We assume C$(b)$ is in excess, so it is not included in calculating the rate of the reaction.
Therefore, the rate of an Eley-Rideal reaction is given by $k_{er}[G][G(s)]$, where the rate coefficient is
\begin{equation}
    k_{er} = \frac{F_G}{B}\gamma\exp{\left(
    -\frac{E}{RT}\right)}.
\end{equation}
Theoretically, $\gamma \in [0,1]$, but the ACA model experimentally sets the $\gamma$-values to match molecular beam data.
Therefore, that constraint is relaxed to $\gamma \in [0,\infty)$, and $\gamma > 1$ signifies that the reaction is not an elementary reaction, which is a well-defined, single-step reaction resulting from a single collision between two or rarely three molecules or ions.
Instead, there are intermediate reactions untracked by the model, which potentially increase uncertainty in the model.
Note, reaction $15$ in \Cref{tab:prata} is an Eley-Rideal reaction, but $\gamma = 10^8$ is a catch-all constant for $\frac{F_G}{B}\gamma$ instead of strictly being a pre-exponential factor.
Therefore, the resulting rate constant is $k_{15} =10^8\exp{\left(-\frac{20676}{T}\right)}$.

\firstline{
Langmuir-Hinshelwood reactions describe surface reactions between two adsorbed species that also contribute to surface carbon removal and gas-phase product formation.}
The rate of a Langmuir-Hinshelwood reaction is given by $k_{lh}[G(s)][G(s)]$, where the rate coefficient is

\begin{equation}
    k_{lh} = \sqrt{\frac{A_v}{B}}F_{G,2D}\gamma\exp{\left(
    -\frac{E}{RT}\right)},
\end{equation}
and
\begin{equation*}
    F_{G,2D} = \sqrt{\frac{\pi k_b T}{2 m_G}}
\end{equation*}
is the mean thermal speed of the adsorbed species on the surface.

\firstline{
The ACA model combines adsorption, desorption, Eley-Rideal, and Langmuir-Hinshelwood reactions to define a high-fidelity finite-rate gas-surface chemistry model.}
These 20 reactions define a system of ordinary differential equations (ODEs) that govern the time evolution of adsorbed species and gas-phase products. The resulting production rates of the ACA model for each species are:

\begin{align}
    \frac{d[\text{O}(s)]}{dt} &= k_{1}[\text{O}][(s)] - k_{2}[\text{O}(s)] - k_{3}[\text{O}][\text{O}(s)] - k_{4}[\text{O}][\text{O}(s)] \nonumber \\
    &\quad\quad  - 2k_{9}[\text{O}(s)]^2 + 2k_{16}[\text{O}_2][(s)]^2 - k_{17}[\text{O}_2][\text{O}(s)] - k_{18}[\text{O}_2][\text{O}(s)] \nonumber \\[.75em]
    \frac{d[\text{O}^*(s)]}{dt} &= k_{5}[\text{O}][(s)] - k_{6}[\text{O}^*(s)] - k_{7}[\text{O}][\text{O}^*(s)] - 2k_{8}[\text{O}^*(s)]^2 \nonumber \\
    &\quad\quad + 2k_{19}[\text{O}_2][(s)]^2 - k_{20}[\text{O}_2][\text{O}^*(s)] \nonumber \\[.75em]
    f_{\text{CO}} =\frac{d[\text{CO}]}{dt} &= k_{3}[\text{O}][\text{O}(s)] + k_{7}[\text{O}][\text{O}^*(s)] + k_{17}[\text{O}_2][\text{O}(s)] + k_{20}[\text{O}_2][\text{O}^*(s)] \nonumber \\[.75em]
   f_{\text{CO}_2} = \frac{d[\text{CO}_2]}{dt} &= k_{4}[\text{O}][\text{O}(s)] + k_{18}[\text{O}_2][\text{O}(s)] \nonumber \\[.75em]
    f_{\text{O}} =\frac{d[\text{O}]}{dt} &= \frac{P_\text{O}}{A_v \sqrt{2 \pi m_{\text{O}} k_b T}}
    - k_{1}[\text{O}][(s)] + k_{2}[\text{O}(s)] - k_{4}[\text{O}][\text{O}(s)] \nonumber \\ 
    &\quad\quad  - k_{5}[\text{O}][(s)] + k_{6}[\text{O}^*(s)] + k_{18}[\text{O}_2][\text{O}(s)] \label{species_odes}  \\[.75em]
    f_{\text{O}_2} =\frac{d[\text{O}_2]}{dt} &= \frac{P_{\text{O}_2}}{A_v \sqrt{2 \pi m_{\text{O}_2} k_b T}}
    + k_{8}[\text{O}^*(s)]^2 + k_{9}[\text{O}(s)]^2 \nonumber \\
    &\quad\quad - k_{16}[\text{O}_2][(s)]^2 - k_{18}[\text{O}_2][\text{O}(s)] - k_{19}[\text{O}_2][(s)]^2 \nonumber 
        \end{align}
    \begin{align}
    \frac{d[\text{N}(s)]}{dt} &= k_{10}[\text{N}][(s)] - k_{11}[\text{N}(s)] - k_{12}[\text{N}][\text{N}(s)] - k_{13}[\text{N}][\text{N}(s)] \nonumber \\
    &\quad\quad - 2k_{14}[\text{N}(s)]^2 - k_{15}[\text{N}(s)] \nonumber \\[.75em]
    f_{\text{CN}} =\frac{d[\text{CN}]}{dt} &=  k_{12}[\text{N}][\text{N}(s)] + k_{15}[\text{N}(s)] \nonumber \\[.75em]
    f_{\text{N}} =\frac{d[\text{N}]}{dt} &= \frac{P_{\text{N}}}{A_v \sqrt{2 \pi m_{\text{N}} k_b T}} - k_{10}[\text{N}][(s)] + k_{11}[\text{N}(s)] - k_{13}[\text{N}][\text{N}(s)] \nonumber \\[.75em]
    f_{\text{N}_2} =\frac{d[\text{N}_2]}{dt} &= k_{13}[\text{N}][\text{N}(s)] + k_{14}[\text{N}(s)]^2. \nonumber
\end{align}
Note, the change in concentration of a gas-phase species is also referred to as the flux (e.g., $f_{\text{CO}} =\frac{d[\text{CO}]}{dt}$).
Surface-bound species such as O$(s)$, O$^*(s)$, and N$(s)$, are not assigned fluxes in this context.
The total site density of the surface $B$ is held constant resulting in
\begin{equation*}
\begin{aligned}
    B &= [\text{O}(s)] + [\text{O}^*(s)] + [\text{N}(s)] + [(s)].
\end{aligned}
\end{equation*}

\firstline{
In simulations using the ACA model, surface coverage is assumed to reach a steady state at each flow solver timestep.}
This steady state is found by setting the rate of change in surface coverage equal to zero (i.e., $\frac{d[\text{O(s)}]}{dt} = \frac{d[\text{O}^*\text{(s)}]}{dt} = \frac{d[\text{N(s)}]}{dt} = 0$).
In a large-scale CFD simulation of a hypersonic flow environment, the surface coverage is calculated at each flow solver timestep, representing the fraction of available surface sites occupied by adsorbed gas species.
Assuming the ACA model is at a steady state does not imply chemical equilibrium: it implies that the local surface coverage adapts instantly \cite{prata2022air}.
However, the surface chemistry continues changing over the course of the CFD simulation.

\firstline{
To compute steady-state surface coverages in the ACA model, we solve a nonlinear equation for the empty site concentration $[(s)]$ using a closed-form formulation of the surface species.}
We first define intermediate quantities  that compactly represent combinations of reaction rate constants and gas-phase concentrations:
{\allowdisplaybreaks
\begin{align*}
    A_1 &= 2k_{16}[\text{O}_2] \\
    B_1 &= k_{1}[\text{O}] \\
    C_1 &= 2k_{9} \\
    D_1 &= k_{2} + \left(k_{3} + k_{4} \right)[\text{O}] + \left(k_{17} + k_{18} \right)[\text{O}_2] \\
    A_2 &= 2k_{19}[\text{O}_2] \\
    B_2 &= k_{5}[\text{O}] \\
    C_2 &= 2k_{8} \\
    D_2 &= k_{6} + k_{7}[\text{O}] + k_{20}[\text{O}_2] \\
    A_3 &= 0 \\
    B_3 &= k_{10}[\text{N}] \\
    C_3 &= 2k_{14} \\
    D_3 &= k_{11} + k_{15}+ \left(k_{12} + k_{13} \right)[\text{N}].
\end{align*}
}
These constants are used to express the steady-state surface coverages of weakly bonded oxygen $[\text{O}(s)]$, strongly bonded oxygen $[\text{O}^*(s)]$, and bonded nitrogen $[\text{N}(s)]$ in terms of the unknown free site concentration $[(s)]$ via:
\begin{align}
[\text{O}(s)] &= \frac{2\left(A_1[(s)]^2 + B_1[(s)]\right)}{D_1 + \sqrt{D_1^2 + 4C_1 (A_1[(s)]^2 + B_1[(s)])}}, \label{hifi_weak} \\
[\text{O}^*(s)] &= \frac{2\left(A_2[(s)]^2 + B_2[(s)]\right)}{D_2 + \sqrt{D_2^2 + 4C_2 (A_2[(s)]^2 + B_2[(s)])}}, \label{hifi_strong} \\
[\text{N}(s)] &= \frac{2 B_3[(s)]}{D_3 + \sqrt{D_3^2 + 4C_3 B_3[(s)]}}. \label{hifi_n}
\end{align}
The total active site density 
$B$ is constant, and $[(s)]$ is found by enforcing the conservation constraint:
\begin{equation} \label{hifi_s}
\begin{aligned}
\relax [(s)] = B &- \frac{2\left(A_1[(s)]^2 + B_1[(s)]\right)}{D_1 + \sqrt{D_1^2 + 4C_1 (A_1[(s)]^2 + B_1[(s)])}} \\
&- \frac{2\left(A_2[(s)]^2 + B_2[(s)]\right)}{D_2 + \sqrt{D_2^2 + 4C_2 (A_2[(s)]^2 + B_2[(s)])}} - \frac{2 B_3[(s)]}{D_3 + \sqrt{D_3^2 + 4C_3 B_3[(s)]}}.
\end{aligned}
\end{equation}
We solve for $[(s)]$ using the bisection root-finding algorithm \cite{2020SciPy-NMeth} with bounds
$[(s)]\in[0, B]$. 
Then, that root is substituted 
into \Crefrange{hifi_weak}{hifi_n} to obtain 
the surface coverage of each bonding group.
Finally, the steady-state fluxes can be solved by substituting the surface coverages into \Cref{species_odes}.
\subsection{Low-Fidelity Model}\label{lofi}
\firstline{
The low-fidelity model simplifies the high-fidelity ACA model by omitting particular reactions.}
We remove nitrogen-related reactions since they only indirectly affect CO production.
We also remove the reactions involving weakly-bonded O since it is expected to contribute less to CO production.
Recall the adsorption selectivity is larger in reaction five compared to reaction one in \Cref{tab:prata}, and the pre-exponential factor in reaction 20 is ten times larger than in reaction 17.
The resulting reduced ACA model is shown in \Cref{tab:red_prata}, which involves six reaction mechanisms. The low-fidelity model describes only oxidation by O and O$_2$ and surface catalyzed recombination of O.
This type of model reduction—omitting species or pathways deemed less influential—follows standard practice in chemical kinetics to enable more efficient analysis while retaining dominant behavior \cite{steinfeld1999chemical}.

\begin{table}[H]
    \caption{ Reduced ACA gas-surface chemistry model reaction set}
    \centering
    \begin{tabular}{clccl}
                \hline
                Number   & Reaction & $S$ or $\gamma$ &   $E/R$ & Type    \\\hline
                5  & O + (s)  $\rightarrow$ O$^*$(s)  & 0.7  &  0      & Adsorption  \\
                6  & O$^*$(s) $\rightarrow$ O + (s)  & 1.0  & 96500   & Desorption  \\
                7  & O + O$^*$(s) + C(b) $\rightarrow $ CO + O + (s)  & 1000 & 4000 & Eley-Rideal  \\
                8  & O$^*$(s) +  O$^*$(s) $\rightarrow$ O$_2$ + 2(s) & $10\textsuperscript{-3}$  & 15000 & Langmuir-Hinshelwood  \\
                19  & O$_2$ + 2(s) $\rightarrow$ 2O$^*$(s)  & 1.0 &  8000 & Adsorption  \\
                20  & O$_2$ + O$^*$(s) + C(b)  $\rightarrow$ CO + O$_2$ + (s) & 1000.0  &  4000      & Eley-Rideal  \\
                \hline
    \end{tabular}
    \label{tab:red_prata}
\end{table}

The resulting production rates for each species are
\begin{equation}
\begin{aligned}
    \frac{d[\text{O}^*(s)]}{dt} &= k_{5}[\text{O}][(s)] - k_{6}[\text{O}^*(s)] - k_{7}[\text{O}][\text{O}^*(s)] - 2k_{8}[\text{O}^*(s)]^2 \\
    &\quad + 2k_{19}[\text{O}_2][(s)]^2 - k_{20}[\text{O}_2][\text{O}^*(s)]  \\[.75em]
    f_{\text{CO}} = \frac{d[\text{CO}]}{dt} &= k_{7}[\text{O}][\text{O}^*(s)] + k_{20}[\text{O}_2][\text{O}^*(s)] \\[.75em]
    f_{\text{O}} = \frac{d[\text{O}]}{dt} &= \frac{P_\text{O}}{A_v \sqrt{2 \pi m_{\text{O}} k_b T}}
    - k_{5}[\text{O}][(s)] + k_{6}[\text{O}^*(s)] \\[.75em]
    f_{\text{O}_2} = \frac{d[\text{O}_2]}{dt} &= \frac{P_{\text{O}_2}}{A_v \sqrt{2 \pi m_{\text{O}_2} k_b T}}
    + k_{8}[\text{O}^*(s)]^2 
    - k_{19}[\text{O}_2][(s)]^2.
\end{aligned}
\end{equation}
The total site density of the surface is
\begin{equation}
\begin{aligned}
    B &= [\text{O}^*(s)] + [(s)].
\end{aligned}
\end{equation}
A steady state solution to the reduced ACA model is found by setting the rate of change in surface coverage equal to zero (i.e., $ \frac{d[\text{O}^*\text{(s)}]}{dt} = 0$).
Following the steady-state solution of the ACA model, let
\begin{align*}
    A_2 &= 2k_{19}[\text{O}_2] \\
    B_2 &= k_{5}[\text{O}] \\
    C_2 &= 2k_{8} \\
    D_2 &= k_{6} + k_{7}[\text{O}] + k_{20}[\text{O}_2].
\end{align*}
Then
\begin{equation}
    [O^*(s)] = \frac{2\left(A_2[(s)]^2 + B_2[(s)]\right)}{D_2 + \sqrt{D_2^2 
    + 4C_2 \left(A_2[(s)]^2 + B_2[(s)] \right)}}
\end{equation}
and
\begin{equation}
\begin{aligned}
    \relax[(s)] = B - \frac{2\left(A_2[(s)]^2 + B_2[(s)]\right)}{D_2 + \sqrt{D_2^2 
    + 4C_2 \left(A_2[(s)]^2 + B_2[(s)] \right)}}.
\end{aligned}
\end{equation}

\firstline{
The low-fidelity model achieves a substantial reduction in complexity by limiting the reaction set and species considered.}
Specifically, the high-fidelity ACA model consists of $20$ reactions and $11$ species, whereas the low-fidelity model retains only six key reactions and five species by focusing on a reduced set of species primarily involving oxygen and its strongly-bonded surface forms.
This reduction preserves the dominant oxidation and recombination mechanisms relevant to CO production while improving model tractability.
\subsection{Inputs and the Quantity of Interest}\label{input_qoi}
\begin{figure}[H]
    \centering
    \includegraphics[width=0.4\textwidth]{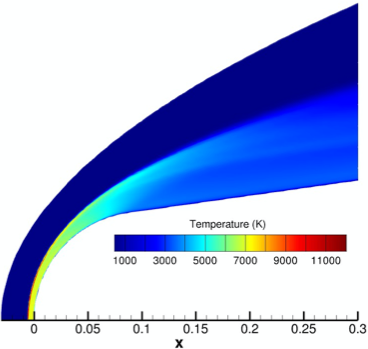}
    \caption[Flow field temperature contour]{Temperature contour of the flow field around the sphere-cone geometry ($x$ in m) for freestream velocity of 7 km/s at an altitude of $40$ km. Image from \cite{mussoni2022sensitivity}.}
    \label{fig:spherecone}
\end{figure}

\firstline{
To generate inputs of a simulated planetary re-entry vehicle's hypersonic flight, we use a simulation test case from Mussoni et al.~\cite{mussoni2022sensitivity} that was previously  outlined by Candler et al.~\cite{candler2012nonequilibrium, candler2017characterization}.}
Specifically, the planetary re-entry vehicle's geometry is chosen as a $10$ centimeter radius sphere -$8^{\circ}$ cone as shown in \Cref{fig:spherecone}, where $x$ is the longitudinal distance from the stagnation point (i.e., the tip of the nose).
Freestream conditions correspond to altitudes of $(20, 25, 30, 35, 40)$ kilometers with a freestream speed of $7$ kilometers per second.
A chemically reacting laminar boundary layer is applied, and the boundary conditions are in non-ablating, non-catalytic, radiative equilibrium.
Therefore, the gas temperature at surface is equal to the surface temperature.
We use the software US3D \cite{candler2015development} to simulate the CFD.
The US3D data supplies densities of the incoming gas-phase species, total gas pressures, and the temperatures at discretized longitudinal distances from the stagnation point.
Following Candler, we subtract $1000$ Kelvin from each input temperature for a more realistic temperature regime because the boundary condition does not account for all heat losses \cite{candler2012nonequilibrium}.

\firstline{
To prepare inputs for the ablation models, CFD data involving additional gas species are converted into fluxes and molar concentrations for the relevant species.}
The test case involves more gas species than our ablation models, which we represent as $G' = G \cup \{ \text{N}_2, \text{NO} \} = \{ \text{O}, \text{O}_2, \text{N}, \text{N}_2, \text{NO} \}$.
To calculate the ablation models' incoming gas species fluxes and molar concentrations, we transform the simulation densities $\rho_{G'}$ to partial pressures $P_{G'}$:
\begin{equation}
\begin{aligned}
    c_{G'} &= \frac{\rho_{G'}}{M_{G'}} \\
    c_{\text{total}} &= \sum c_{G'} \\
    \chi_{G'} &= \frac{c_{G'}}{c_{\text{total}}} \\
    P_{G'} &= \chi_{G'} P_{\text{total}}.
\end{aligned}
\label{eq:convert}
\end{equation}
Then, partial pressures are used to derive the input flux of the gas-phase species:
\begin{equation}
     f_G = \frac{P_G}{A_v \sqrt{2\pi m_G k_b T}},
     \label{eq:molar_flux}
\end{equation}
and their molar concentrations:
\begin{equation}
     c_G = \frac{P_G}{RT}.
\label{eq:molar_conc}
\end{equation}
Note, the molar concentrations in \Cref{eq:molar_conc} are not the same as \Cref{eq:convert} because temperatures have been shifted by $1000$ Kelvin.
The results from \Cref{eq:molar_flux} and \Cref{eq:molar_conc} are inputs to the ablation models. The total site density of the surface is set as $B=1\times 10^{-5}$.

\begin{figure}[H]
    \centering
    \begin{subfigure}[b]{0.42\textwidth}
         \centering
         \includegraphics[width=\textwidth]{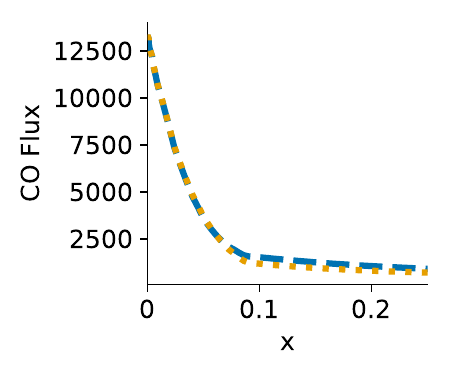}
         \captionsetup{skip=-8pt} 
         \caption{Altitude: 20 km}
         \label{fig:CO_lo_20}
     \end{subfigure}
     \begin{subfigure}[b]{0.52\textwidth}
         \centering
         \includegraphics[width=\textwidth]{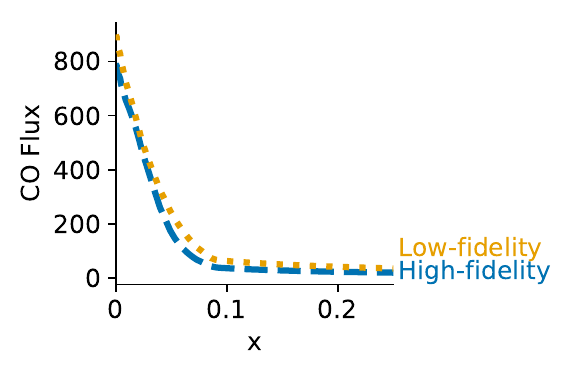}
         \captionsetup{skip=-8pt} 
         \caption{Altitude: 40 km}
         \label{fig:CO_lo_40}
     \end{subfigure}
    \caption{CO flux in mol$\cdot$m$^{-2}\cdot$s$^{-1}$ outputs from the high-fidelity model (blue dashed curve) and low-fidelity model (orange dotted curve) for the (a) 20km and (b) 40km input scenarios.}
    \label{fig:CO_lo}
\end{figure}

\firstline{
In this study, our primary goal is to quantify the recession rate of the carbon surface, which we estimate from the amount of CO being produced in the Eley-Rideal and Langmuir-Hinshelwood reactions.}
Omitting reactions in the low-fidelity model has two distinct effects on the CO flux. 
First, it eliminates certain pathways to CO production, which leads to an underestimate of the high-fidelity ACA model's CO flux, as shown in \Cref{fig:CO_lo_20}. 
Second, it reduces competition for free surface sites, which leads to an overestimate of the high-fidelity ACA model's CO flux, as shown in \Cref{fig:CO_lo_40}. Since the surface recession rate is determined by the total flux of carbon species relative to the bulk carbon material, and CO is the dominant product in oxidation environments, variations in the CO flux directly reflect changes in the predicted recession rate.

\begin{figure}[H]
    \centering
     \includegraphics[scale=0.95]
     {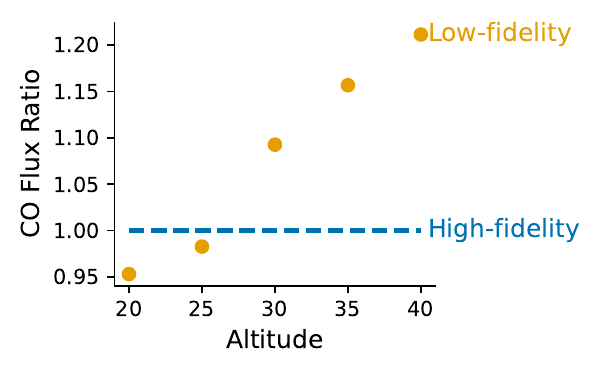}
    \captionsetup{skip=-6pt} 
    \caption{Comparison of the CO flux ratio for each altitude.}
    \label{fig:CO_ratio}
\end{figure}

\firstline{
To assess model accuracy in predicting CO flux, we define a QoI called the CO flux ratio.}
For a given altitude, the cumulative CO flux discrepancy between the low- and high-fidelity model is:
\begin{equation}
    \text{CO flux ratio} = 
    R(\text{low}) = \frac{\sum_{i=1}^d f_{\text{CO}}^{(\text{low})}(x_i)}{\sum_{i=1}^d f_{\text{CO}}^{(\text{hi})}(x_i) }, 
\end{equation}
where $d=72$ discrete spatial points along the arc length from $x=0$ to $x=0.25$ meters.
Depending on the altitude, the low-fidelity model can under- or overpredict the cumulative CO flux, as shown in \Cref{fig:CO_ratio}.
The low-fidelity model can overpredict by more than $20\%$ in the $40$km scenario.
Additionally, underpredicting CO flux creates a serious risk: it can lead to selecting a heat shield that is too thin to safely return a planetary re-entry vehicle to Earth when used in TPS design.
Therefore, we aim to improve the low-fidelity model's accuracy in both under- and overprediction scenarios.

\section{Model Enrichment Approach}\label{enrichment}
\begin{figure}[H]
\tikzstyle{terminator} = [rectangle, draw, text centered, rounded corners, minimum height=2em, text width=5.0cm]
\tikzstyle{process} = [rectangle, draw, text centered, text width=4.0cm, minimum height=2em]
\tikzstyle{data}=[trapezium, draw, text centered, trapezium left angle=60, trapezium right angle=120, minimum height=1.5em, text width=3.5cm]
\tikzstyle{connector} = [draw, -latex']
\centering
\begin{tikzpicture}[node distance=2cm and 6cm]

\node [terminator,fill=navyblue!50] at (0,2.25) (C) {\textbf{High-fidelity model} \\ 20 reactions; 11 species};
\node [terminator, fill=orange] at (0,0) (R) {\textbf{Low-fidelity model} \\ 6 reactions; 5 species};
\node [terminator, fill=black!20] at (0,-3) (E) {\textbf{Enriched model} \\ 9 reactions; 7 species};

\node [data, fill=navyblue!50] at (6,2.25) (data) {Calibration data};
\node [process, fill=black!20] at (6,-3) (disc) {\textbf{Data-driven enrichment}};

\path [connector] (C) -- (data);
\path [connector] (C) -- (R);
\path [connector] (R) -- (E);
\path [connector] (data) -- (disc);
\path [connector] (disc) -- (E);

\node[draw=none, text width=3.5cm] at (1.85,1.15) (omit) {Omit reactions};
\node[draw=none, text width=4.0cm] at (2.1,-1.5) (embed) {Embed physics-based \\ and data-driven \\ enrichments};
\node[draw=none, text width=3.2cm] at (7.7,0.0) (cali) {Calibrate \\ data-driven \\ enrichment};

\end{tikzpicture}
\caption{Schematic of the high-fidelity, low-fidelity, and enriched models. The low-fidelity model omits some reactions, and the enriched model incorporates physics-based and data-driven corrections, calibrated using high-fidelity observations.}
\label{fig:model_diagram}
\end{figure}
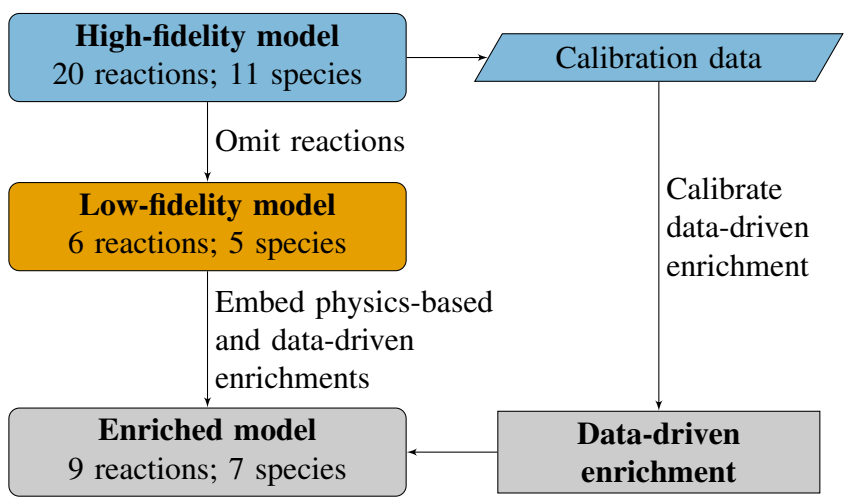

\firstline{
To reduce the discrepancy in the CO flux ratio, we add enrichments to the low-fidelity model.}
In the high-fidelity model, nitrogen-related reactions and weakly-bonded oxygen reactions are only coupled to the strongly-bonded oxygen reactions through their mutual interactions with the surface; they are competing for free surface sites. In the low-fidelity model, the only type of surface interaction is oxygen bonding strongly to the surface, which can result in a surplus of CO production.
A deficit in CO production can also occur from eliminating the weakly-bonded oxygen Eley-Rideal reaction depending on input conditions.
Usually, we see underproduction in high temperature regimes and overproduction at lower temperatures.
To address the low-fidelity model discrepancy, we form an enriched model by adding physics-based and data-driven enrichments to the low-fidelity model, depicted in \Cref{fig:model_diagram}.

\subsection{Physics-Based Enrichments}
\begin{figure}[H]
    \centering
    \begin{subfigure}[b]{0.3\textwidth}
         \centering
         \includegraphics[width=\textwidth]{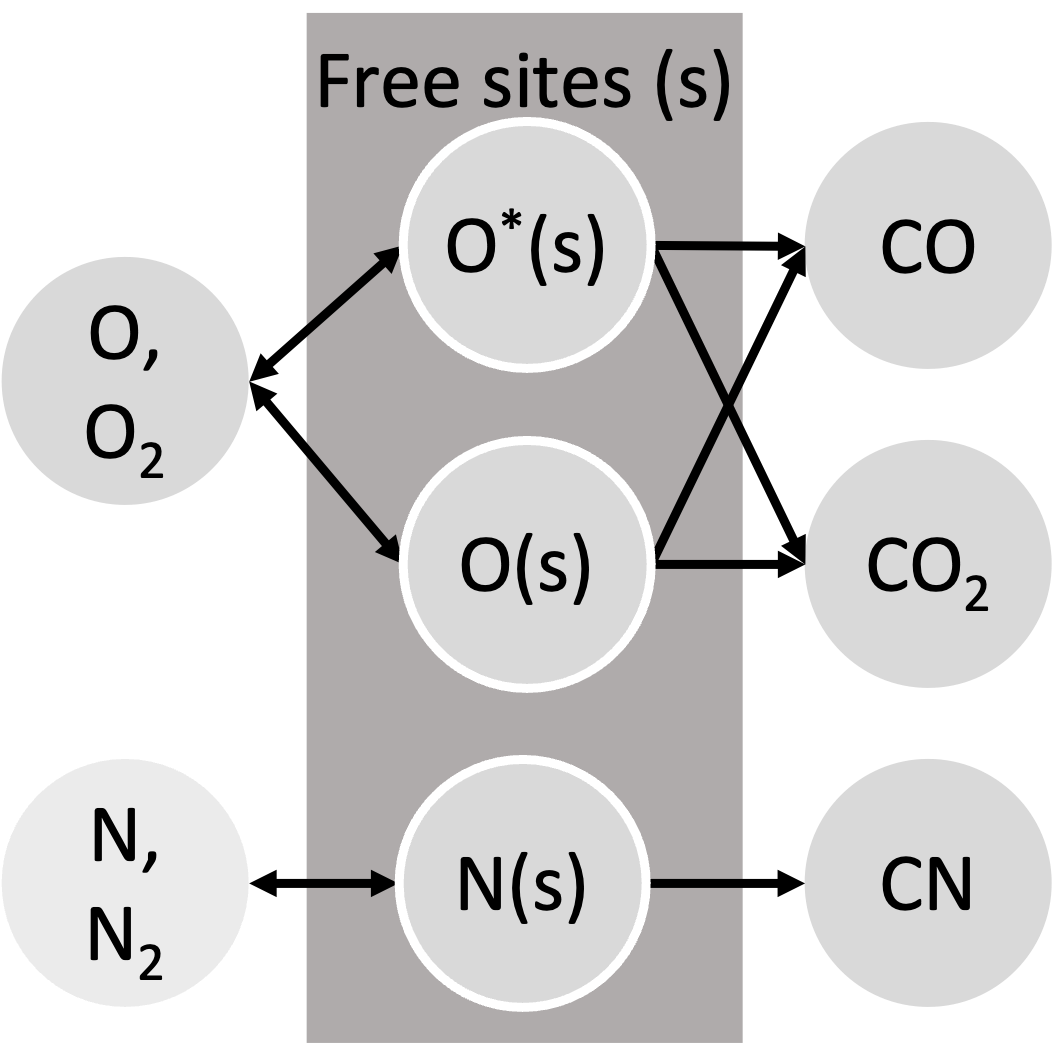}
         \caption{High-fidelity model}
     \end{subfigure}
     \hspace{0.02\textwidth}
     \begin{subfigure}[b]{0.3\textwidth}
         \centering
         \includegraphics[width=\textwidth]{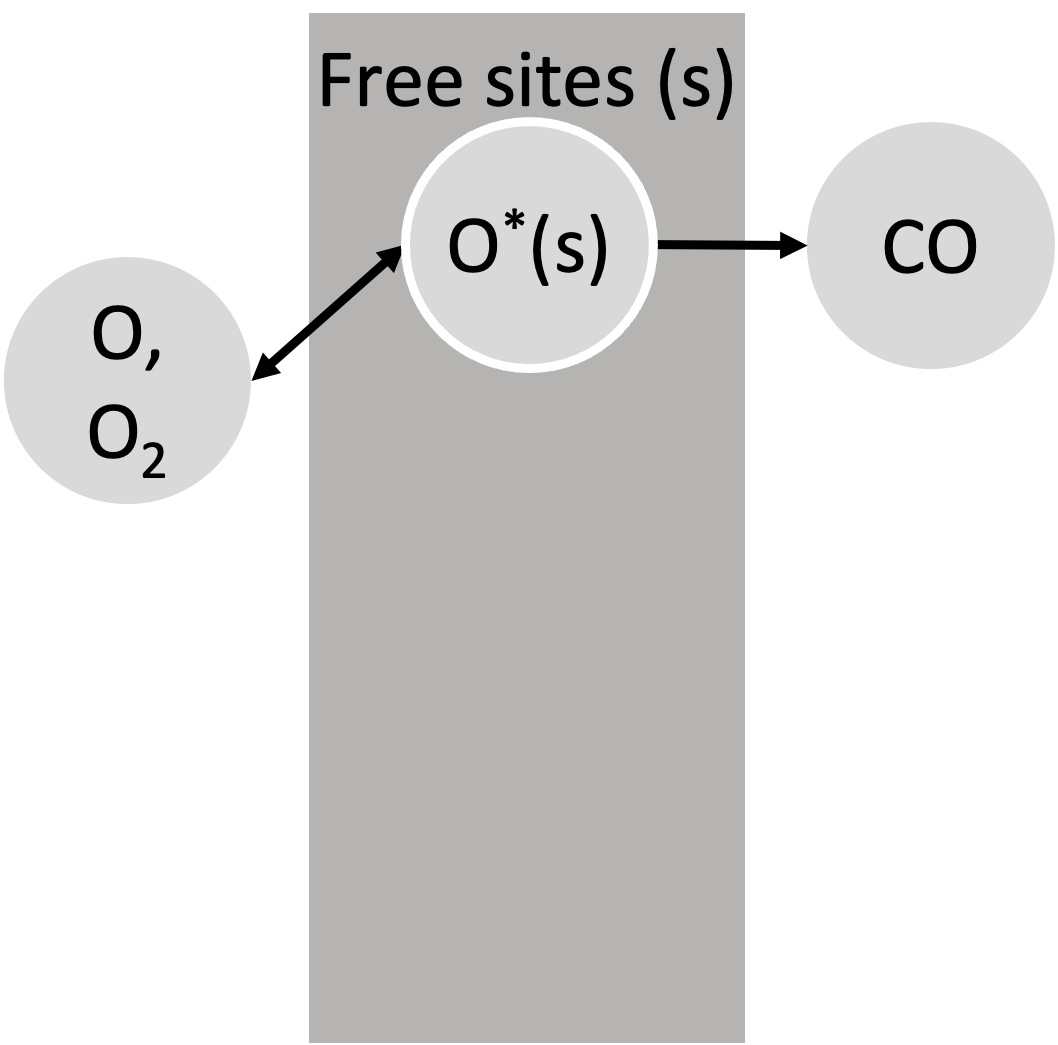}
         \caption{Low-fidelity model}
     \end{subfigure}
     \hspace{0.02\textwidth}
     \begin{subfigure}[b]{0.3\textwidth}
         \centering
         \includegraphics[width=\textwidth]{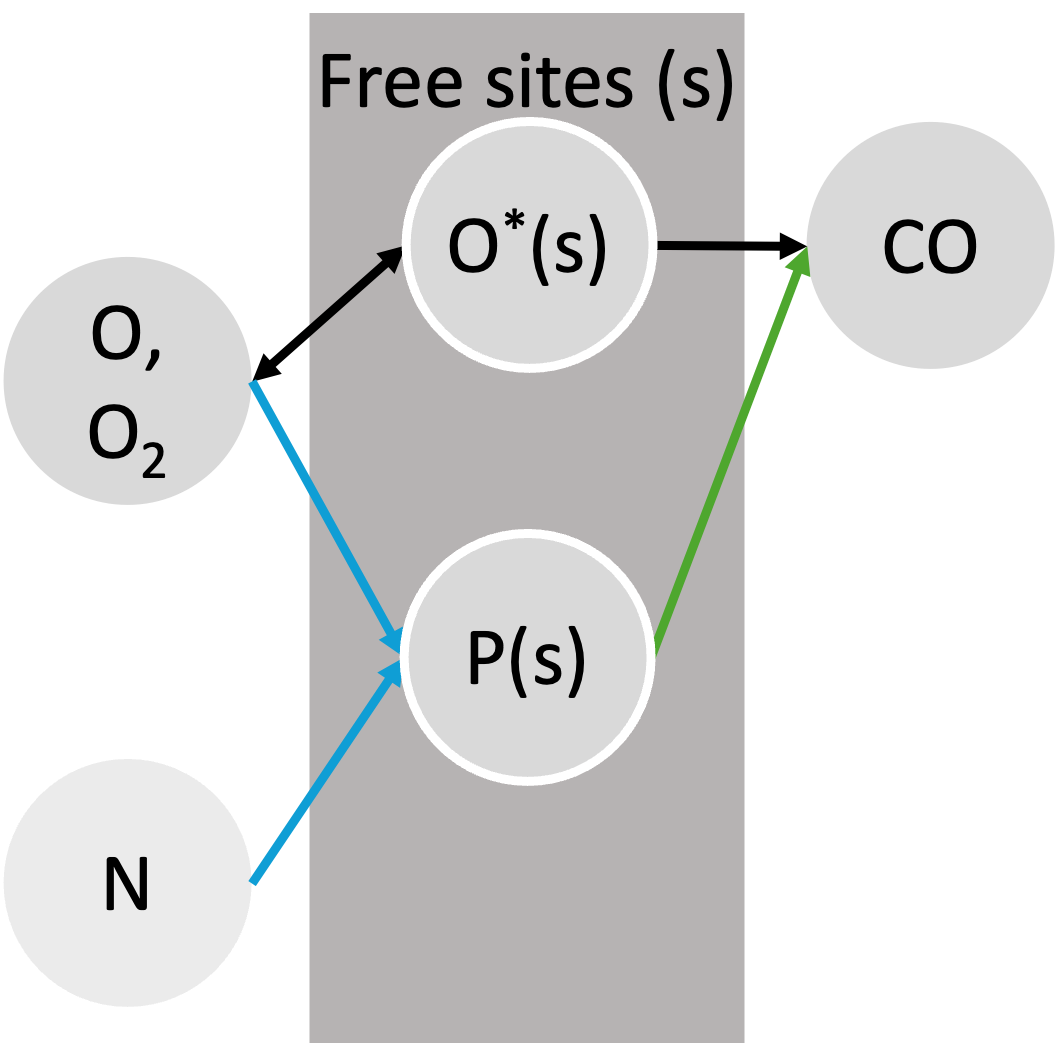}
         \caption{Enriched model}
     \end{subfigure}
     \vspace{-6pt}
    \caption{Graphical representations of (a) high-fidelity, (b) low-fidelity, and (c) enriched models. Blue arrows show physics-based adsorption rates and the green arrow data-driven pseudo-rate.}
    \label{fig:model_drawings}
\end{figure}

\firstline{
To account for both sources of error—unmodeled surface competition and unmodeled CO production—we add three enrichments represented as three reactions to the low-fidelity model, as shown in \Cref{fig:model_drawings}.}
To allow for surface competition, we add two reactions to the low-fidelity model.
Specifically, they are adsorption reactions of oxygen and nitrogen creating a ``placeholder" species bonded to the surface, P(s). 
The placeholder species does not differentiate between the gas species reactants that formed it; it is a catch-all compartment representing all unmodeled surface bonding. 
To allow for unmodeled CO production, we add a reaction to the low-fidelity model, namely, a pseudo-reaction producing CO and a free surface site from the placeholder bonded to the surface. The enriched model's complete reaction set is show in \Cref{tab:enrich}.

\begin{table}[H]
    \caption{ Enriched gas-surface chemistry model reaction set}
    \centering
    \begin{tabular}{clccl}
                \hline
                Number   & Reaction & $S$ or $\gamma$ &   $E/R$ & Type    \\\hline
                5  & O + (s)  $\rightarrow$ O$^*$(s)  & 0.7  &  0      & Adsorption  \\
                6  & O$^*$(s) $\rightarrow$ O + (s)  & 1.0  & 96500   & Desorption  \\
                7  & O + O$^*$(s) + C(b) $\rightarrow $ CO + O + (s)  & 1000 & 4000 & Eley-Rideal  \\
                8  & O$^*$(s) +  O$^*$(s) $\rightarrow$ O$_2$ + 2(s) & $10\textsuperscript{-3}$  & 15000 & Langmuir-Hinshelwood  \\
                19  & O$_2$ + 2(s) $\rightarrow$ 2O$^*$(s)  & 1.0 &  8000 & Adsorption  \\
                20  & O$_2$ + O$^*$(s) + C(b)  $\rightarrow$ CO + O$_2$ + (s) & 1000.0  &  4000      & Eley-Rideal  \\
                \blue{$1p$}  & \blue{O + (s)  $\rightarrow$  P(s)}   & \blue{0.3} &  \blue{0}      & \blue{Adsorption} \\
                \blue{$2p$}  & \blue{N + (s) $\rightarrow$ P(s)} & \blue{1.0} &  \blue{2500}   & \blue{Adsorption} \\
                \blue{$3p$}  & \blue{ P(s) + C(b)  $\rightarrow$  CO + (s)}   & \blue{-} &  \blue{-}      & \blue{Pseudo} \\
                \hline
    \end{tabular}
    \label{tab:enrich}
\end{table}

\firstline{
The reactions for placeholder adsorption are grounded in the physics of the high-fidelity ACA model \cite{prata2022air}.}
Specifically, the reactants in placeholder reactions $1p$ and $2p$ mirror those in ACA reactions $1$ and $10$, respectively.
Since they are strictly forward reactions, we use their reaction rates directly.

\firstline{
The enriched model modifies the species production rates and surface site balance by incorporating the effects of the added placeholder reactions.}
The augmented production rates for each species in the resulting enriched model are
\begin{equation}
\begin{aligned}
    \frac{d[\text{O}^*(s)]}{dt} &= k_{5}[\text{O}][(s)] - k_{6}[\text{O}^*(s)] - k_{7}[\text{O}][\text{O}^*(s)] - 2k_{8}[\text{O}^*(s)]^2 \\
    &\quad + 2k_{19}[\text{O}_2][(s)]^2 - k_{20}[\text{O}_2][\text{O}^*(s)]  \\[.75em]
    \blue{\frac{d[\text{P}(s)]}{dt}} &\blue{= k_{1p}[\text{O}][(s)] + k_{2p}[\text{O}][(s)] - k_{3p}[\text{P}(s)]} \\
    f_{\text{CO}} = \frac{d[\text{CO}]}{dt} &= k_{7}[\text{O}][\text{O}^*(s)] + k_{20}[\text{O}_2][\text{O}^*(s)] \blue{+ k_{3p}[\text{P}(s)]}\\[.75em]
    f_{\text{O}} = \frac{d[\text{O}]}{dt} &= \frac{P_\text{O}}{A_v \sqrt{2 \pi m_{\text{O}} k_b T}}
    - k_{5}[\text{O}][(s)] + k_{6}[\text{O}^*(s)] \blue{-k_{p1}[\text{O}][(s)]} \\[.75em]
    f_{\text{O}_2} = \frac{d[\text{O}_2]}{dt} &= \frac{P_{\text{O}_2}}{A_v \sqrt{2 \pi m_{\text{O}_2} k_b T}}
    + k_{8}[\text{O}^*(s)]^2 
    - k_{19}[\text{O}_2][(s)]^2 \\
    \blue{f_{\text{N}} = \frac{d[\text{N}]}{dt}} &\blue{= k_{2p}[\text{O}][(s)]}.
\end{aligned}
\end{equation}
The total site density of the surface is
\begin{equation}
\begin{aligned}
    B &= [\text{O}^*(s)] + [(s)] \blue{+ [\text{P}(s)]}.
\end{aligned}
\end{equation}
A steady state solution to the reduced ACA model is found by setting the rate of change in surface coverage equal to zero (i.e., $ \frac{d[\text{O}^*\text{(s)}]}{dt} = \frac{d[\text{P}\text{(s)}]}{dt} = 0$).
Following the steady-state solution of the low-fidelity model, let
\begin{align*}
    A_2 &= 2k_{19}[\text{O}_2] \\
    B_2 &= k_{5}[\text{O}] \\
    C_2 &= 2k_{8} \\
    D_2 &= k_{6} + k_{7}[\text{O}] + k_{20}[\text{O}_2] \\
    \blue{A_4} &\blue{= 0 }\\
    \blue{B_4} &\blue{= k_{1p}[\text{O}] + k_{2p}[\text{N}] } \\
    \blue{C_4} &\blue{= 0 } \\
    \blue{D_4} &\blue{= k_{3p}}.
\end{align*}
Then
\begin{equation}
    [\text{O}^*\text{(s)}] = \frac{2\left(A_2[(s)]^2 + B_2[(s)]\right)}{D_2 + \sqrt{D_2^2 
    + 4C_2 \left(A_2[(s)]^2 + B_2[(s)] \right)}},
\end{equation}
\begin{equation}
    \blue{[\text{P}(s)] = \frac{B_2[(s)]}{D_2}},
\end{equation}
and
\begin{equation}
\begin{aligned}
    \relax[(s)] = B - \frac{2\left(A_2[(s)]^2 + B_2[(s)]\right)}{D_2 + \sqrt{D_2^2 
    + 4C_2 \left(A_2[(s)]^2 + B_2[(s)] \right)}} - \blue{\frac{B_2[(s)]}{D_2}}.
\end{aligned}
\end{equation}

\firstline{
The enriched model enhances the low-fidelity formulation by introducing critical physical mechanisms while maintaining a manageable level of complexity.}
It includes nine reactions and seven species, bridging the gap between the low- and high-fidelity models.
By incorporating reactions that account for surface site competition and additional CO production pathways, the enriched model improves physical realism without reverting to the full complexity of the high-fidelity model.

\subsection{Pointwise Enrichment}
\firstline{
The third enrichment is a pseudo-reaction that does not correspond to a physical reaction in the ACA model.}
Unlike the placeholder adsorption enrichments, this reaction lacks a direct analog in the high-fidelity chemistry and is thus referred to as a pseudo-reaction.
It is not a realistic recombination reaction because the bonded placeholder species and bulk carbon would not be the only reactants needed for the bond to break.
Additionally, the placeholder species does not differentiate between O and N in P$(s)$, but the pseudo-reaction produces CO, violating conservation of atoms. 
It might be possible to mitigate this effect by tracking mixture fractions of O and N in P$(s)$, but this would likely increase the computational complexity of the enrichment and is thus left to future work if necessary.
With no further physics-based information to impose on the reaction, we now aim to extract its functional form from data.
In this case we use simulation data from the high-fidelity model to inform the enrichment, but in general experimental data could be used.

\firstline{
To formulate the pseudo-reaction, we first extract pointwise reaction rates by fitting the enrichment to high-fidelity CO flux data.}
Lacking intuition about the appropriate functional dependencies of the pseudo-reaction, we adopt a staged, data-driven approach.
First, we fit a pointwise reaction rate at each calibration data point, indexed by altitude and spatial position $(a, x)$, where the model inputs are defined as $\bm{\xi}_{a,x} = (a, x, T, P_{\text{total}}, \bm{f}_G, \bm{c}_G)$.
These inputs, derived from the CFD simulation, include temperature, total pressure, incoming species fluxes, and incoming molecular concentrations, all of which vary with altitude and location.
The pointwise loss is defined as the negative log-likelihood of observing the high-fidelity CO flux from the enriched model with a given $k_{3p}$:
\begin{equation}
    L(k_{3p};\bm{\xi}_{a,x}) \equiv -\log l(\log(k_{3p});\bm{\xi}_{a,x}) \propto \frac{\left(f_{\text{CO}}^{(hi)}(\bm{\xi}_{a,x})-f_{\text{CO}}^{(en)}(\bm{\xi}_{a,x};\log(k_{3p})))\right)^2}{2\sigma^2},
    \label{eq:loss_pointwise}
\end{equation}
where $\sigma = 0.05f_{\text{CO}}^{(hi)}(\bm{\xi}_{a,x})$.
Minimizing this loss yields a pointwise optimal value of the pseudo-reaction rate constant:
\begin{equation}
    k_{3p}^*(\bm{\xi}_{a,x}) = \min_{k_{3p}} L(k_{3p}; \bm{\xi}_{a,x}).
    \label{eq:opt_pointwise}
\end{equation}
We solve the minimization problem~\Cref{eq:opt_pointwise} using the Nelder-Mead optimization method~\cite{2020SciPy-NMeth}, producing a set of input–output pairs $(\bm{\xi}_{a,x},  k_{3p}^*)$. These results form the foundation for identifying a functional form for $k_{3p}$ in our data-driven enrichment.

\firstline{
Beyond enabling construction of a data-driven functional form, the pointwise enrichment provides a direct, interpretable representation of the reaction behavior across relevant input space.}
By computing the optimal reaction rates 
$k_{3p}^*(\bm{\xi}_{a,x})$ at each calibration point, we obtain a set of empirical values that can be used to assess the validity of proposed physics-based enrichments, such as an Arrhenius equation.
These proposed enrichments can be evaluated by comparing their predictions to the pointwise values, revealing where they align and where they fail.
In this way, the pointwise enrichment serves not only as a method for data fitting but also as a diagnostic tool for model selection and validation, offering a means to test physical assumptions directly against data informed behavior without assuming a specific functional form in advance.

\subsection{Data-Driven Enrichment}
\firstline{
Given the pointwise enrichment defined above, we build a data-driven enrichment.}
Pointwise estimates are used to train a Gaussian process (GP), which serves as a nonparametric representation for predicting $\log(k_{3p})$ with associated uncertainty.
GPs offer several advantages in this setting: they provide well-calibrated uncertainty estimates, are analytically tractable, and scale reasonably for problems with low input-output dimensionality and limited data \cite{williams2006gaussian}.
While more flexible models like Bayesian neural networks or neural networks with conformal prediction could be used, GPs are well suited for this application due to their computational efficiency and robustness for small datasets.
Alternatively, interpretable models like SINDy could also be employed, and the pointwise estimates enables direct evaluation of such models' suitability by comparing their predictions to the empirically derived rates.

\firstline{
The inputs to the GP are drawn from the set of ablation model inputs $\bm{\xi}_{a,x}$.}
However, we do not anticipate the pseudo-reaction rate depends on all of these inputs.
To identify a parsimonious and effective input subset, we apply the Least Absolute Shrinkage and Selection Operator (LASSO) for feature selection \cite{scikit-learn}.
The most influential features are then used as inputs to train the GP model.

\firstline{
Our GP includes a nugget term to account for observational noise and model misspecification \cite{gramacy2012cases}.}
The nugget allows for the possibility that the selected inputs do not fully explain the variability in $\log(k_{3p})$, reflecting epistemic uncertainty stemming from unmodeled phenomena.
This feature is crucial for maintaining a well-calibrated uncertainty estimate, particularly in regions where the training data may be sparse or the functional relationship may be more complex than what the selected inputs can capture.

\firstline{
In summary, this data-driven enrichment enhances the low-fidelity model by learning a functional approximation of the pseudo-reaction rate with quantified uncertainty.}
This enables closer alignment with high-fidelity observations and provides a foundation for more accurate predictions in the results that follow.
\section{Numerical Results}\label{results}
\firstline{This section presents the numerical results of our modeling and enrichment framework.}
The results are organized into three parts: Pointwise Enrichment Results, Data-Driven Enrichment Results, and Enriched Model Results.

\subsection{Pointwise Enrichment Results}
\begin{figure}[H]
    \centering
    \begin{subfigure}[b]{0.42\textwidth}
         \centering
         \includegraphics[width=\textwidth]{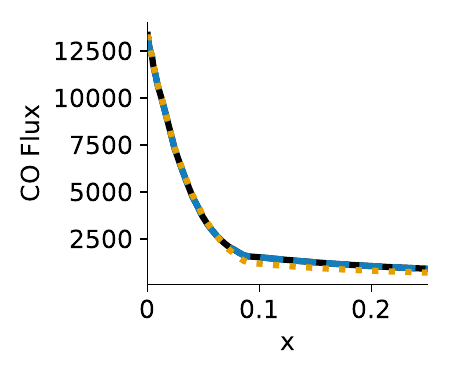}
         \captionsetup{skip=-8pt} 
         \caption{Altitude: 20 km}
         \label{fig:CO_pointwise_20}
     \end{subfigure}
     \begin{subfigure}[b]{0.52\textwidth}
         \centering
         \includegraphics[width=\textwidth]{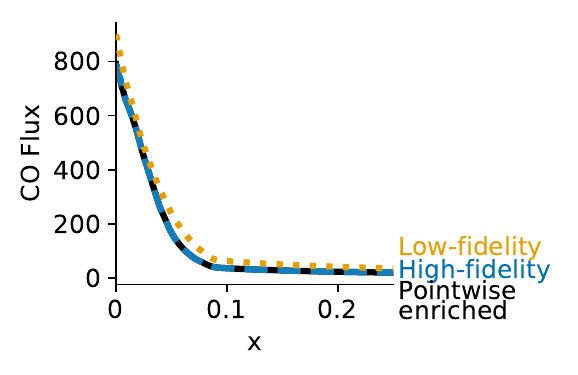}
         \captionsetup{skip=-8pt}
         \caption{Altitude: 40 km}
         \label{fig:CO_pointwise_40}
     \end{subfigure}
     \vspace{-6pt}
    \caption{CO flux outputs from the high-fidelity model (blue dashed curve), low-fidelity model (orange dotted curve), and pointwise enriched model (black curve) for the (a) 20km and (b) 40km input scenarios.}
    \label{fig:CO_pointwise}
\end{figure}

\firstline{
The pointwise enriched model exactly reproduces the high-fidelity CO flux at calibration points, but it is not deployable without further generalization.}
As shown in \Cref{fig:CO_pointwise}, the enriched model with pointwise values of $k_{3p}$ can perfectly match the high-fidelity calibration data.
However, computing a new pointwise enrichment at an unseen altitude–spatial location pair ($a$,$x$) would require evaluating the high-fidelity model at that point, so this is not a practically-deployable enrichment as-is.
Thus, in the second stage of our data-driven enrichment formulation,  we use the collection of pointwise enrichments obtained from calibration data to identify a functional form for $k_{3p}$ that generalizes across inputs.

\begin{figure}[H]
    \centering
    \includegraphics[scale=0.95]{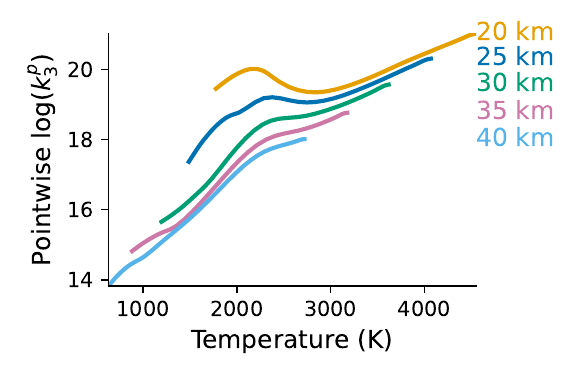}
    \captionsetup{skip=-6pt} 
    \caption{Pointwise rates for the pseudo-reaction for the five altitude scenarios.}
    \label{fig:pointwise_v_temp}
\end{figure}

\firstline{
We use pointwise enrichment results to evaluate whether standard chemical rate expressions, such as the Arrhenius form, are sufficient to represent the pseudo-reaction.}
Since the Arrhenius equation depends solely on temperature, we can visualize the pointwise $k_{3p}$ values vs.~surface temperature and observe whether the pointwise $k_{3p}$ yields a functional dependence on temperature.
As shown in~\Cref{fig:pointwise_v_temp}, the spread in pointwise $\log(k_{3p})$ values across altitudes for the same surface temperature indicates the pseudo-reaction does not depend on temperature alone.
Thus, no Arrhenius-type reaction depending on temperature alone can capture the functional behavior of the pseudo-reaction. Having ruled out the most common chemical reaction form, we turn to a nonparametric approach to identify the functional form of the  pseudo-reaction.

\subsection{Data-Driven Enrichment Results}
\firstline{
We calibrate the data-driven enrichment using scenarios at
$20$, $30$, and $40$ km and reserve the $25$ and $35$ km scenarios for validation.}
Because pointwise $\log(k_{p3})$ exhibits a positive, linear trend with respect to the input features, we first remove this trend to allow the GP to focus on modeling the more complex residual structure.
To do this, we fit a linear regressor to the pointwise values of $\log(k_{p3})$, using $\log(\text{temperature})$, $\log(\text{total pressure})$, and the molecular concentration of N as inputs from the calibration data.
Then, we compute the residuals from the linear regressor, which serve as the training data for fitting the GP.

\firstline{
Our Gaussian process model uses a combined Radial Basis Function (RBF) kernel and a white noise kernel to capture the structure and noise in the data.} 
The RBF kernel is selected for its ability to model smooth, continuous functions by assuming that inputs closer in feature space produce more similar outputs, which is an assumption consistent with the behavior of the pseudo-reaction rate.
To model unresolved errors and observational noise in the calibration data, we include a white noise kernel with a fixed noise level of 0.005.
This noise level was selected to ensure that all calibration data points fall within the $95\%$ confidence interval of the GP predictions, providing a conservative estimate of uncertainty while avoiding overfitting.
Only the RBF lengthscales, initialized to one, are optimized during training to preserve robustness and allow the GP to adapt to key trends.
The combination of the two kernels results in a flexible yet stable representation that captures both systematic behavior and epistemic uncertainty.

\begin{figure}[H]
    \centering
    \includegraphics[scale=0.88]{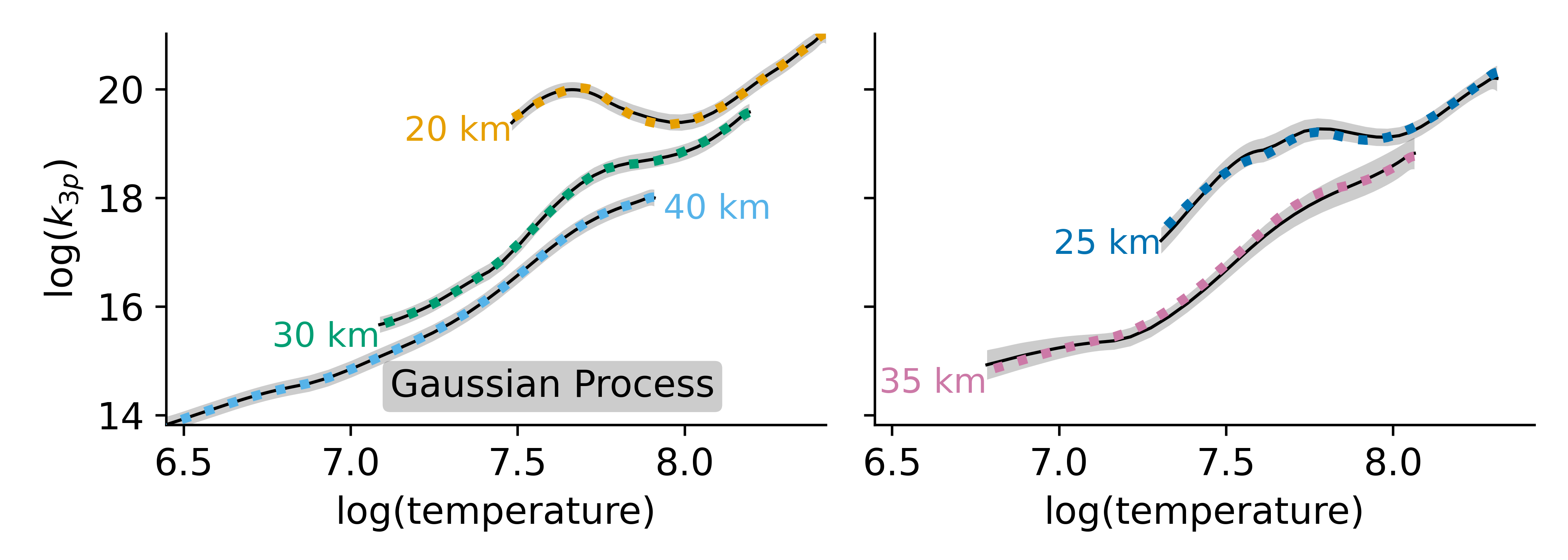}

    \vspace{-8pt}  

    \makebox[0.48\linewidth][c]{(a) Calibration Scenarios}
    \hfill
    \makebox[0.48\linewidth][c]{(b) Validation Scenarios}
    \vspace{-6pt}
    \caption{Pointwise enrichment (colored dots) versus a Gaussian process with mean (black curve) and $95\%$ confidence interval (gray shaded region) for (a) calibration and (b) validation scenarios.}
    \label{fig:kp3_GP}
\end{figure}

\firstline{
The performance of our trained GP is evaluated by comparing its predictions against the pointwise enrichment for both the calibration and validation scenarios as shown in \Cref{fig:kp3_GP}.}
The calibration data, which were used to fit the GP, demonstrate a strong alignment with the GP predictions, indicating that our data-driven enrichment effectively captures the underlying trends in the data.
Furthermore, the validation data also falls within the $95\%$ confidence interval of the GP predictions.
This result underscores the robustness of our data-driven enrichment, as it not only generalizes well to unseen data but also maintains a high level of predictive certainty across the range of inputs.
The ability of the GP to encompass all validation data points within its $95\%$ confidence interval supports its effectiveness as a predictive tool in this context; however, we note that the validation scenarios are interpolative, falling within the range of training data, and do not test the model's extrapolative performance.

\subsection{Enriched Model Results}
\begin{figure}[H]
    \centering
    \includegraphics[scale=0.95]{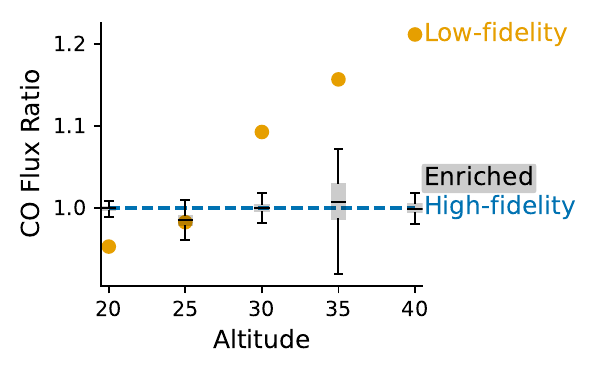}
    \captionsetup{skip=-6pt} 
    \caption{Comparison of the CO flux ratio from the low-fidelity model (orange dots) and probabilistic enriched model (box plots) for each altitude.
    Each box shows the interquartile range (IQR), with the median marked by the center line and whiskers extending to the most extreme values within $1.5\times$ IQR.}
    \label{fig:CO_ratio_enriched}
\end{figure}

\firstline{
To quantify uncertainty in the enriched model, we propagate samples from the data-driven pseudo-reaction rate to predictions of the CO flux ratio.}
After qualitatively validating the data-driven pseudo-reaction rate, we embed it and the physics-based adsorption rates back into the low-fidelity model to form the enriched model.
The resulting enriched model is probabilistic because the pseudo-reaction rate is a GP.
Therefore, we sample $100$ realizations for the pseudo-reaction rate and propagate them forward to the QoI:
\begin{equation}
    \text{CO flux ratio} = 
    R(\text{en}) = \frac{\sum_{i=1}^d f_{\text{CO}}^{(\text{en})}(x_i)}{\sum_{i=1}^d f_{\text{CO}}^{(\text{hi})}(x_i) }, 
\end{equation}
where $d=72$ discrete spatial points along the arc length from $x=0$ to $x=0.25$ meters.
Results are shown in \Cref{fig:CO_ratio_enriched}, where the enriched model's QoI is represented as a box plot for each altitude. 

\firstline{
In our results, the box plots illustrate the uncertainty captured by the enriched model.}
This uncertainty stems from the probabilistic nature of the GP used to model the pseudo-reaction rate, combined with the deterministic physics-informed reactions.
Unlike the deterministic low-fidelity model, which yields a single prediction, the enriched model provides a distribution of CO flux ratios across the GP samples.
This comparison demonstrates the enhanced capability of the enriched model to represent the complexities of the system, as it not only captures the high-fidelity CO flux ratios but also provides associated uncertainties, which are critical for informed decision-making.

\section{Conclusion}\label{conc}
\firstline{
In this work, we developed a hybrid enriched ablation model that integrates physics-based and data-driven enrichments to improve low-fidelity model predictions of the QoI.}
Motivated by the need for fast yet reliable tools for heat shield design, the low-fidelity model omits nitrogen-related surface competition and weakly-bonded oxygen reactions.
To mitigate the errors caused by those omissions, we introduced three placeholder surface reactions: two to restore surface competition and one to compensate for missing CO production.
Physics and the high-fidelity model informed the reaction mechanisms of the three placeholder reactions as well as the reaction rates of the two adsorption placeholder reactions.
We formulated the data-driven enrichment using a staged approach, beginning with pointwise fits to high-fidelity data and finishing with a GP model for the pseudo-reaction rate.

\firstline{
By embedding hybrid enrichments into the low-fidelity model, we produced an enriched model that captures key trends in the high-fidelity QoI and supplies meaningful uncertainty estimates.}
The enriched model improves predictive accuracy across altitudes and provides a probabilistic framework suitable for downstream applications in design and risk analysis.

\firstline{
This study demonstrates the utility of combining physical insight with statistical learning to enhance low-fidelity models.}
Future work will focus on integrating the enriched model into online CFD simulations.
To manage computational cost, we will explore a staged approach, decoupling simulation and training the data-driven enrichment.
We also plan to investigate how the model’s uncertainty estimates can be used to detect out-of-distribution behavior.
Finally, we aim to extend the enriched model to handle multiple interacting sources of model-form error simultaneously with multiple data-driven enrichments.

\section*{Acknowledgments}
This work was supported by the Laboratory Directed Research and Development (LDRD) program at Sandia National Laboratories,
a multimission laboratory managed and operated by National Technology \& Engineering Solutions of Sandia, LLC, a wholly owned subsidiary of Honeywell International Inc., for the U.S. Department of Energy’s National Nuclear Security Administration under contract DE-NA0003525. 
LDRD $\#233072$.
This paper describes objective technical results and analysis. Any subjective views or opinions that might be expressed in the paper do not necessarily represent the views of the U.S. Department of Energy or the United States Government. The authors would like to thank Lincoln Collins, JP Heinzen, Kathryn Maupin, and Kyle Neal for many helpful discussions.

\bibliographystyle{IEEEtran}
\bibliography{IEEEabrv,refs}

\end{document}